\documentclass[superscriptaddress,reprint,showpacs,amsmath,amssymb,aps,prl,floatfix]{revtex4-1}

\usepackage{bbold}
\usepackage{graphicx}
\usepackage{dcolumn}
\usepackage{bm}
\usepackage{epstopdf}
\usepackage{ulem}
\usepackage{color}

\newcommand{\er}{Er$^{3+}$}

\newcommand{\eu}{Eu$^{3+}$}
\newcommand{\fdz}{$^5$D$_{0}$}
\newcommand{\sfz}{$^7$F$_{0}$}

\newcommand{\YSO}{Y$_2$SiO$_5$}

\newcommand{\CWO}{CaWO$_4$}

\begin{document}


\title{Stark Echo Modulation  for Quantum Memories}

\author{A. Arcangeli}
\affiliation{NEST, Scuola Normale Superiore and Instituto Nanoscienze-CNR, P.za San Silvestro 12, 56127 Pisa,  Italy.}
\affiliation{PSL Research University, Chimie ParisTech, CNRS, Institut de Recherche de Chimie Paris, 75005, Paris, France}
\author{A. Ferrier}
\affiliation{PSL Research University, Chimie ParisTech, CNRS, Institut de Recherche de Chimie Paris, 75005, Paris, France}
\affiliation{Sorbonne Universit\'es, UPMC Univ Paris 06, Paris 75005, France}
\author{Ph. Goldner}
\email{philippe.goldner@chimie-paristech.fr}
\affiliation{PSL Research University, Chimie ParisTech, CNRS, Institut de Recherche de Chimie Paris, 75005, Paris, France}

\date{\today}

\begin{abstract}
Quantum memories for optical and microwave photons provide key functionalities  in quantum processing and communications. Here we propose a protocol well adapted to solid state ensemble based memories coupled to cavities. It is called Stark Echo Modulation Memory (SEMM), and allows large storage bandwidths and low noise. This is achieved in a echo like sequence combined with phase shifts induced by small electric fields through the linear Stark effect. 
We investigated  the protocol for rare earth nuclear spins   and found 
a high  suppression of unwanted collective emissions that is compatible with single photon level operation. Broadband storage together with  high fidelity for the Stark retrieval process is also demonstrated.
SEMM could be used to store optical or microwave photons in ions and/or spins. This includes NV centers in diamond and rare earth doped crystals, which are among the most promising solid-state quantum memories. 
\end{abstract}

\maketitle

Quantum memories (QM) are essential components in quantum information processing.  They enable storage and on-demand retrieval of quantum states  and allow using  fast but short-lived processing qubits, or photonic states that are excellent  carriers of quantum information but are difficult to store. QM for light find applications in linear optics quantum computing, as well as in quantum communications and networks, where they could enable distribution of entangled states over long distances using quantum repeater architectures \cite{Kimble:2008if,Northup:2014gv}.  There is also a growing interest in spin based quantum memories that store micro-wave photons which in turn can be interfaced to superconducting qubits  \cite{Kurizki:2015vy}.
In the solid state,  optical and microwave QM based on inhomogeneously broadened ensemble are actively investigated in rare earth (RE) ion doped crystals and  diamonds containing NV centers \cite{Bussieres:2014dc,Saglamyurek:2015esa,Goldner:2015ve,Wolfowicz:2015ex,Probst:2015ku,Kubo:2011dxa}
These two systems are 
well adapted to highly multimode storage, where multiple photons with large bandwidths are stored for long times \cite{Afzelius:2009gc}. Moreover, high efficiency can be obtained by coupling these centers to a cavity, overcoming their weak  interactions with photons, either for spin or optical transitions \cite{Afzelius:2010je,Afzelius:2013ga,Julsgaard:2013br}.
A natural protocol to implement QM in inhomogeneous  ensembles is the spin or photon echo \cite{Hahn:1950ge,Abella:1966vb} which recovers the initial excitation of the system by applying a $\pi$ pulse to the storage transition. This inverts the atomic or spin phase evolution and results in a collective emission, the so-called echo.
However, this scheme does not allow low-noise operation, a key parameter for quantum memories, which  must store photonic qubits like single photons \cite{Ruggiero:2009uu}. This is because the collective emission occurs in an inverted medium which produces a too large spontaneous emission at the memory output. To avoid this situation, several protocols  have been proposed and experimentally investigated.
However, they require spectral tailoring \cite{Nilsson:2005ea,Alexander:2006it,deRiedmatten:2008ck,McAuslan:2011ke,Chaneliere:2015cd}, which requires a long lived storage level and can reduce bandwidth,  or particular spatial phase matching conditions \cite{Damon:2011tx}, that are difficult to combine with a cavity. Another possibility is to use fast frequency tunable cavities \cite{Afzelius:2013ga,Julsgaard:2013br}, that may be technologically challenging for micro-wave  high Q cavities or in the  the optical domain.  

Here we propose and experimentally investigate a  protocol, inspired by the Stark echo modulation spectroscopic technique \cite{Mims:1964cj}, in which small electric fields are used to shift the phase of subgroups of  ions or spins in a sequence with two $\pi$ pulses. This allows controlling the collective emissions, without spectral tailoring or spatial phase matching, and in fixed frequency cavities with medium finesse. The Stark Echo Modulation Memory (SEMM) protocol is therefore particularly relevant for 
ensembles of RE or NV spins coupled to superconducting resonators  \cite{Afzelius:2013ga,Julsgaard:2013br}. It could also by used in \er doped materials to provide a  highly efficient cavity-enhanced memory at the 1.5 $\mu$m telecom wavelength, despite the inefficient spectral tailoring found in these systems \cite{HastingsSimon:2008fj}.  
 In the following, we first show that SEMM is well adapted to broadband and low-noise operation. We then report on experimental investigations in an ensemble of RE nuclear spins, confirming our analysis and demonstrating a  $\approx 10^{-5}$ suppression in intensity of the intermediate collective emission, and a 99.9 \% average fidelity of the Stark retrieval process determined by quantum state tomography.

\begin{figure}
\includegraphics[width=0.85\columnwidth]{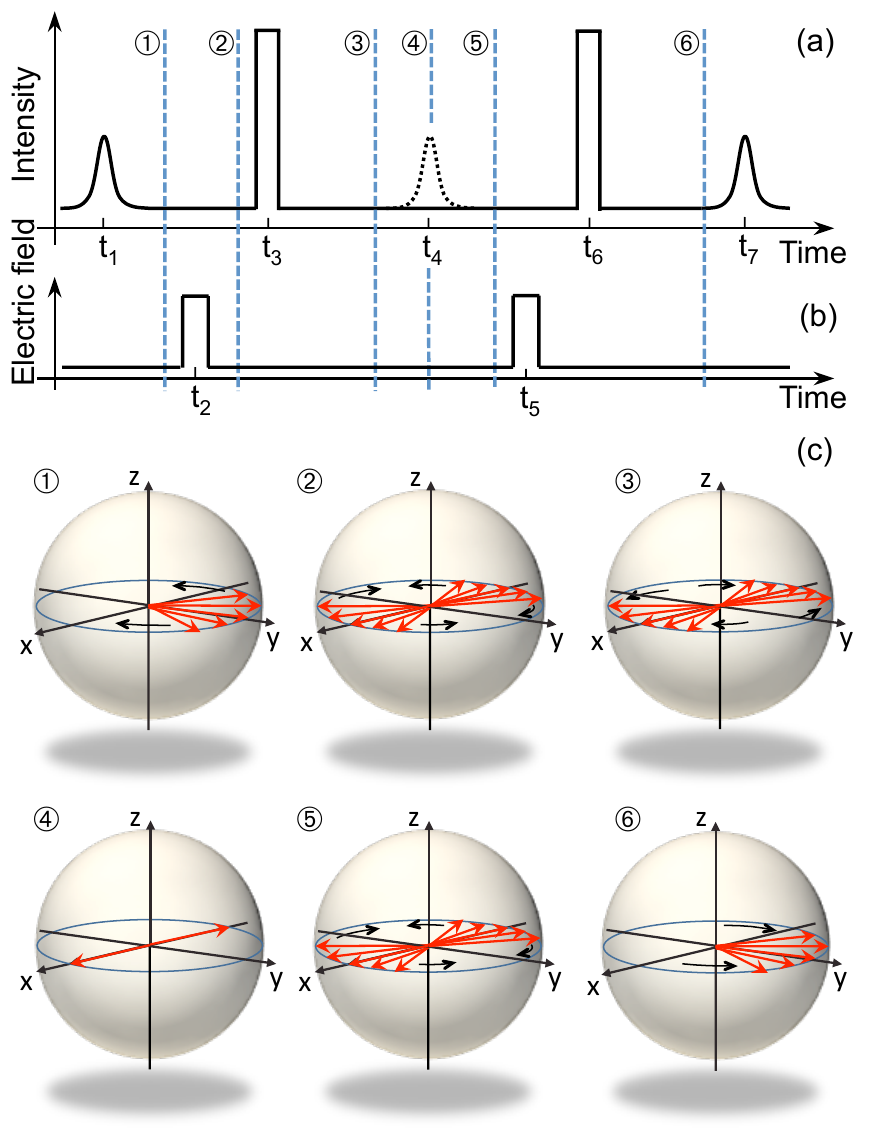}
\caption{The SEMM sequence. (a) microwave or optical fields. The memory input is at $t_1$, $\pi$ pulses at $t_3$ and $t_6$, and the output at $t_7$. (b) electric field. The Stark pulses at $t_2$ and $t_5$  produce a phase shift that cancels the collective emission at $t_4$. (c) Bloch sphere representations of the  wavepackets evolution at the points labeled in (a) and (b). For clarity, the input pulse has an $\pi/2$ area.}
\label{scheme}
\end{figure}

We consider an ensemble of centers in a crystal with an inhomogeneously broadened optical or spin transition showing a linear Stark effect. The ensemble has an inversion symmetry, that can be intrinsic to the host or created by separating the sample in two parts for which the electric field is reversed \cite{Mims:1964cj}. Because of the inversion symmetry, a given electric field will produce a positive frequency shift for half of the centers, and a negative one for the other half.   The SEMM principle is shown in Fig. \ref{scheme}. We assume that the whole sequence takes place within a time  much shorter than the centers' population and coherence lifetimes ($T_1$ and $T_2$, respectively) to preserve a high storage fidelity.  Initially, all centers are in the same state. At time $t_1$, a single input photon is absorbed by the ensemble, and the wavepackets start to dephase relative to each other because of the inhomogeneous broadening. At time $t_2$, an electric field $E$ is applied   to induce a phase shift  $2\pi\Delta = 2\pi E k=\pi/2$ to half of the centers and therefore $-\pi/2$ to the other half, because of the ensemble's inversion symmetry. $k$ is the linear Stark coefficient of one of the subgroups related by the inversion symmetry. The wavepackets divide in two groups with opposite phase shifts, as shown on the Bloch sphere 2 of Fig. \ref{scheme}. At time $t_3$, a 
$\pi$ pulse is applied to the transition,  and for $t>t_3$, the ensemble polarization or magnetization $P(t)$, summed over all centers, is then proportional to:
\begin{equation}
P(t) \propto \int_{-\infty}^{+\infty}e^{i \omega\delta t} \cos \left (2\pi\Delta T_s \right)\mathrm{d} \omega, \label{cancel}
\end{equation}
where $\omega$ is the frequency of the transition (centered at $\omega=0$), $\delta t = 2t_3-t_1-t$ and  $T_s$ is the Stark pulse length. As in a 2-pulse echo experiment, the inhomogeneous broadening is rephased at $t_4=2t_3-t_1$, but $P(t_4)$ vanishes for  $2\pi\Delta T_s=\pi/2$ or $E=1/(4kT_s)$. There is therefore no collective emission (echo) at $t_4$. To recover the input photon from the memory, a second electric field pulse is applied at $t_5$, as well as a second $\pi$ pulse at $t_6$. The polarization at $t>t_6$ is proportional to:
\begin{eqnarray}
P(t) {}\propto{} \left[\int_{-\infty}^{+\infty}e^{i\left(2\pi\Delta T_s- \omega \delta t'-2\pi\Delta T_s\right)}\mathrm{d}  \omega \right.\\ \nonumber
+ \left. \int_{-\infty}^{+\infty}e^{i\left( -2\pi\Delta T_s -\omega \delta t'+2\pi\Delta T_s \right)}\mathrm{d} \omega \right],
\end{eqnarray}
where $\delta t' = 2t_6-t_4-t$.  At $t_7=2t_6-t_4 $, the inhomogeneous broadening is again rephased, whereas the Stark phase shifts cancel, which gives $P(t_7)=P(t_1)$. This collective emission or echo is the output of the memory and is  identical to the initial input  (Fig. \ref{scheme}). Thanks to the two $\pi$ pulses, this emission occurs in a non-inverted medium, which avoids  spontaneous emission at the time and in the mode of the memory output. This is required for the memory to operate in the quantum regime \cite{Ruggiero:2009uu}. 
Another fundamental source of noise is due to spontaneous emission at $t_4$, which would lead to a collective emission at $t_7$, because of the $\pi$ pulse at $t_6$,  with no relation with the memory input \cite{Afzelius:2013ga}. 
This unwanted echo is however cancelled by the $\pm \pi/2$ phase shift produced by the Stark pulse at $t_5$, in the same way as the echo at $t_4$ is suppressed by the Stark pulse at $t_2$ (see Eq. \ref{cancel}). 

Until now, we assumed that the magnitude of the frequency shift induced by the electric field is the same for all centers. However, variations in each center environment will cause a distribution of the Stark coefficients. Moreover, the electric field will also be to some degree spatially inhomogeneous over the sample. This could limit the SEMM to a (small) sub-ensemble of centers.  We examine this question below by considering a distribution of Stark coefficients. Electric field inhomogeneities can be treated in the same way. Assuming  no correlation between the transition broadening $\Gamma$ and the Stark distribution, the polarization after a square electric field pulse of duration $T_s$ and amplitude $E$ is:
\begin{equation}
P = P_0 \int_{-\infty}^{+\infty} \cos(2\pi k E T_s) g(k) d k \label{inhShift}
\end{equation}
where $g$ is the normalized distribution of the Stark coefficients ($\int g(k)\, d k=1$) for one of the subgroups related by the inversion symmetry. 
We have therefore $P=\Re(\tilde{g})$, where  $\tilde{g}$ is the Fourier transform of $g$. $P=0$ will occur for Stark pulse amplitude and duration satisfying:
\begin{equation} 
\Re(\tilde{g}(ET_s))=0.\label{cancelCond}
\end{equation}
In the case, of a symmetric distribution centered on $k_0$, $g(k)=g_1(k-k_0)$, where $g_1(x)=g_1(-x)$.
$P$ is given by: 
\begin{equation}
\Re(\tilde{g}(ET_s))=\cos(2\pi E T_s k_0)\tilde{g}_1(ET_s)
\end{equation}
and cancels for $k_0ET_s=1/4$, i.e. a central phase shift of $\pi/2$, independently of the width of the Stark distribution $g(k)$. As shown in the supplemental material, condition \ref{cancelCond} can be satisfied for any Stark coefficient distribution, unless a large fraction of centers have a zero Stark shift.  
After cancellation of the intermediate echo at $t_4$, the second Stark pulse at $t_5$ is identical to the first one at $t_2$, but induces an opposite phase shift in each wave packet. This results in a complete recovery of the initial input for any distribution $g(k)$.  This would not be the case if an additional $\pi/2$ phase shift was applied, leading to an overall $\pi$ shift, even in the case of a symmetric Stark distribution (see supplemental material).  In SEMM, the memory bandwidth is therefore only limited  by the $\pi$ pulses fidelity over the ensemble of centers. This is in sharp contrast with protocols based on transition broadening  by electric fields \cite{Nilsson:2005ea,Alexander:2006it,deRiedmatten:2008ck,McAuslan:2011ke}, in which the bandwidth is directly dependent  on the magnitude of the Stark shifts that can be induced. 
SEMM has no such limitations, and in the microwave or rf ranges, where $\pi$ pulses of high fidelity and bandwidth can be readily obtained, the entire ensemble inhomogeneous linewidth can be used, as shown in the following. 

\begin{figure*}
\includegraphics[width=0.85\textwidth]{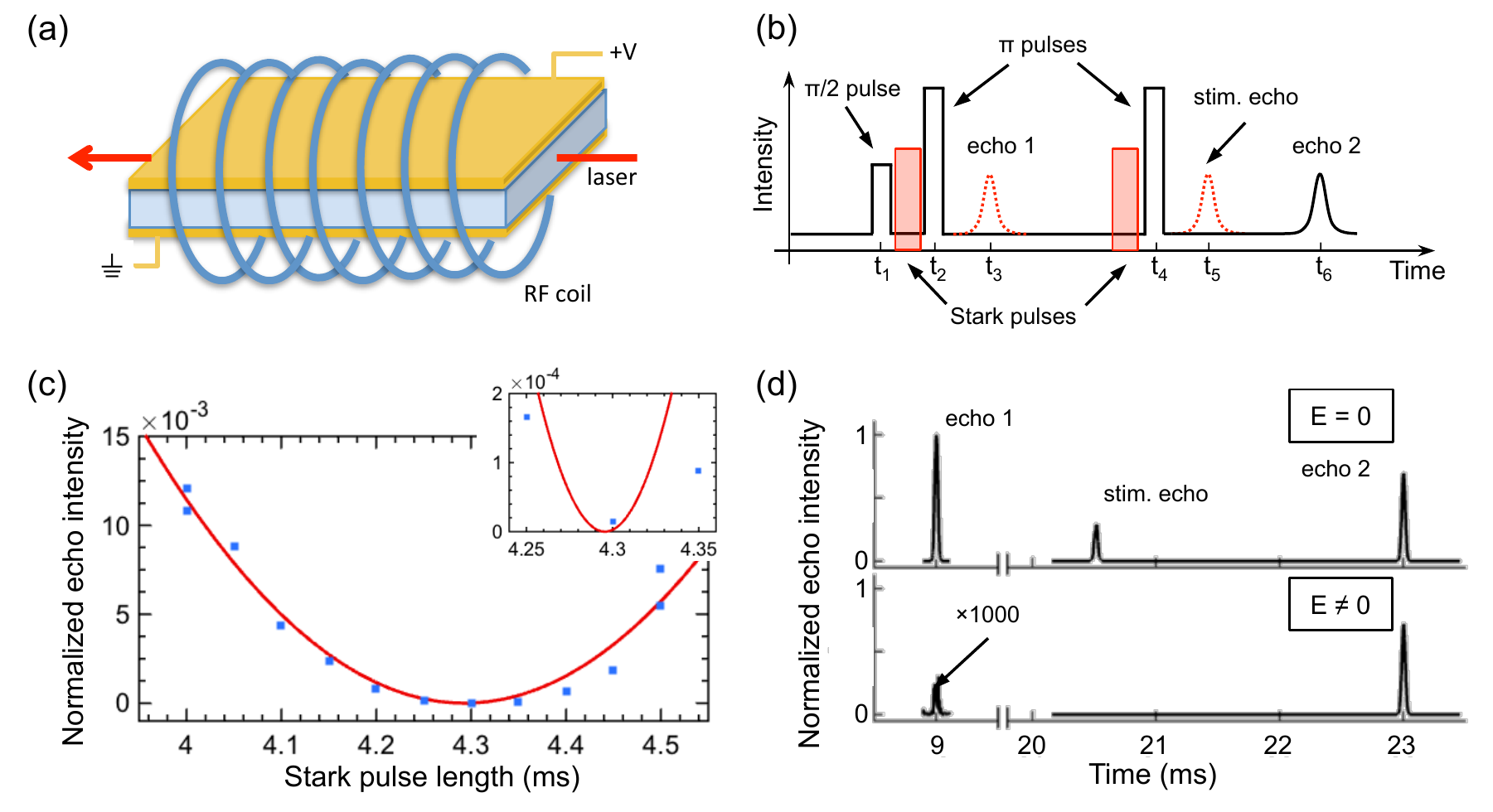}
\caption{\label{exp}Measurements on $^{151}$Eu$^{3+}$:Y$_2$SiO$_5$ nuclear spins. (a) scheme of the sample with attached electrodes to create electric fields. The coil is used to produced rf pulses and the laser to detect spin coherences. (b) Experimental SEMM scheme. Delays were $t_2-t_1=4.5$ ms, $t_4-t_2=11.5 $ ms.  (c) Normalized echo 1 intensity as a function of the length of a  Stark pulse of 16.5 V amplitude. Squares: experimental data; solid line: fit, see text.   (d) Echoes observed with or without electric field in the SEMM sequence.}
\end{figure*}

As a proof of concept, we investigated our protocol in a rare earth doped crystal, \eu:\YSO (Eu:YSO), in which \eu{} ions sit in a $C_1$ symmetry site and the crystal symmetry ($C_{2h}$) includes an inversion operation.  In this material, we recently observed a linear Stark effect on the ground state hyperfine transitions of the $^{151}$\eu{} isotope, which has a nuclear spin $I=5/2$ \cite{Macfarlane:2014fy}. In the present work, rf excitations were stored and retrieved using the ground state  $\pm 1/2 \leftrightarrow \pm 3/2$ transition at 34.58 MHz [$k=0.43$ Hz/(V/cm)], using a 0.1 \% doped sample inserted into a coil [Fig. \ref{exp} (a)]. Spin echoes were optically detected by Raman heterodyne scattering \cite{Mlynek:1983et} using a laser resonant  with the \eu{} \sfz-\fdz{}  transition at 580 nm.  Electric fields parallel to the $D1$ crystal dielectric axis were applied across the 1mm thick sample on which two brass electrodes were placed. All experiments were carried out at 3.5 K. A small static magnetic field of about 48 G was applied in the $D1$ direction to increase the spin coherence lifetime to 25 ms.  Other experimental details can be found in Refs. \onlinecite{Arcangeli:2014dr,Macfarlane:2014fy}. 

The sequence we used is shown in Fig. \ref{exp} (b). We first investigated  suppression of echo 1 after the first rf $\pi$ pulse by applying a Stark pulse  of varying length [Fig. \ref{exp}(c)].     The experimental data, normalized by the echo intensity at zero field, could be well fitted by the equation  $I=(\cos(2\pi \Delta T_s))^2$. The minimum echo intensity corresponds to a suppression $\mu=1.5 \times 10^{-5}$. This was obtained in a sample with no  accurate polishing or parallelism, which is likely to produce inhomogeneous Stark shifts. The observed very low residual echo intensity  therefore confirms the above analysis. 
The lowest achievable echo suppression is limited by parameters fluctuating in time. In our setup, we estimate that the dominating ones were voltage noise, as well as slow fluctuations in temperature and laser intensity and frequency, as signals were averaged  over 200 shots. Echo suppression is particularly important in decreasing the collective emission at the memory output time caused by rephased spontaneous emission (see above). This spontaneous emission can be large when a cavity is used. For example, in a microwave resonator, the Purcell effect and the gain due to the inverted medium result in a number  of spontaneous photons equals to  $n_{sp}=F(e^{Fd}-1)$, where $F$ is the cavity finesse and $d$ the memory opacity \cite{Afzelius:2013ga}. Our experimental value of $\mu$ would allow operation at the single photon level for  a cavity with  $F\approx100$ (see suppl. material). Such a resonator would be suitable  for an impedance matched memory \cite{Afzelius:2013ga} or a strongly coupled one, which has to switch between high and medium finesse to avoid super-radiance during the  microwave pulses \cite{Julsgaard:2013br}.


The complete SEMM sequence was then studied by adding  the second Stark and $\pi$ pulses to retrieve the memory output  [echo 2 in Fig. \ref{exp} (b)]. To optimize the signal to noise ratio, the input of the memory was a $\pi/2$ pulse. The signals recorded at zero electric field are shown in Fig. \ref{exp}(d), upper trace. Besides the intermediate and final echoes, we also observed a stimulated echo after the second $\pi$ pulse. The stimulated echo was separated from the memory output by choosing $t_2-t_1 < (t_4-t_2)/2$. When the Stark pulses were applied, the intermediate echo was strongly suppressed [Fig. \ref{exp} (d)]. The stimulated echo was suppressed too, since it results from a population grating that forms from the pulses at $t_1$ and $t_2$. The second Stark pulse does not induce any additional phase shift on populations and the stimulated echo is suppressed by the first Stark pulse. The memory output, echo 2, is retrieved with an intensity essentially identical to what is observed when no electric field is applied (see below).   The bandwidth of the memory is about 40 kHz limited by the length (24 $\mu$s) of the $\pi$ pulses. This matches well the 32 kHz inhomogeneous width of  the  $\pm 1/2 - 3/2 $ transition at 34.58 MHz \cite{Macfarlane:2014fy}. The length of the memory output pulse was 24 $\mu$s with or without the Stark pulses, showing that SEMM  preserves the full bandwidth, as expected. The frequency shifts due to the Stark field were however only $\pm 58$ Hz, corresponding to  $\approx15$ V applied across 1 mm.
%
%

\begin{figure}
\includegraphics[width=0.8\columnwidth]{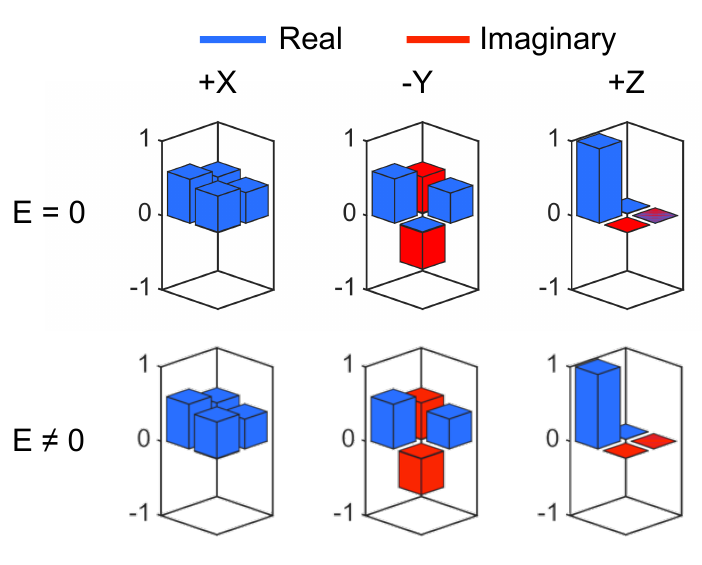}
\caption{SEMM output density matrices with or without electric field in $^{151}$\eu:\YSO{} determined by quantum state tomography. The average fidelity over $\pm X, \pm Y$ and $+Z$ input states is 0.999. }
\label{tom}
\end{figure}

We also performed quantum state tomography to study the influence of the Stark pulses \cite{Morton:2008et}. Input states $\pm X, \pm Y$ were created by varying the phase of the $\pi/2$ pulse, whereas $+ Z$ corresponded to no input pulse. The  $\sigma_X$ and $\sigma_Y$ components of the output density matrix were determined by analyzing the real and imaginary parts of the output pulse. The $\sigma_Z$ component was measured by an additional echo sequence following the output pulse. The upper row of  Fig. \ref{tom} shows the output density matrices for the $+X$, $-Y$ and $+Z$ input states for the SEMM sequence without the Stark pulses. Although the sequence should operate as the identity operation, deviations from the ideal density matrices can be noted and are attributed to phase shifts not exactly compensated in the demodulation circuit and errors induced by the $\pi$ pulses. This however does not prevent the analysis of the Stark pulse effects. The corresponding density matrices  are shown in the bottom row of 
Fig.  \ref{tom}. As can be seen, the density matrices for the three states are nearly not affected by the Stark pulses, resulting in an average  fidelity of 
0.999 for SEMM, taking as a reference the sequence without the Stark pulses. This highlights the robustness of our scheme in which  opposite Stark shifts are used, which compensates phase errors and Stark coefficients distribution.

SEMM could be applied to various  systems. Table \ref{tab} gathers values of $T_2$,  Stark coefficients $k$  and $E_0 = 1/(4kT_2)$ for several optical and spin transitions, assuming $T_s=T_2$ for comparison. In the optical domain, SEMM could be used with rare earth doped doped crystals. As an example, the transition at 580 nm in Eu:YSO exhibits a Stark coefficient of  27 kHz/(V/cm) due to a change in electric dipole moment between ground and excited states. Combined with a coherence  lifetime that can reach 2.6 ms, SEMM would only require  $E_0= 4.6$ mV/cm for a Stark pulse of length $T_2$ (Table \ref{tab}). 
 SEMM could also be used for microwave photons with crystals doped with paramagnetic rare earth ions or diamond containing NV centers. Stark coefficients are much lower since electric fields do not interact directly with spins. However, long $T_2$, which are desirable for memories with long storage time, still allow  electric fields $<$ 10 V/cm to be used for SEMM (Table \ref{tab}).

 \begin{table}
 \caption{Site symmetry, coherence lifetime, Stark coefficient and field for SEMM (assuming $T_s=T_2$) for centers in various hosts with global inversion symmetry.\label{tab}}
 \begin{ruledtabular}
\begin{tabular}{llccc}
System 			& Site  		& $T_2$  						& k 								&$E_{0}$\\
	     			& sym.  		& (ms) 						&	(Hz cm/V)				& (V/cm) \\
Optical trans.			\\
\hspace{0.5 cm}\eu:\YSO			& $C_{1}$ 	&  2.6   \cite{Equall:1994dn} 		& 27000 \cite{Macfarlane:2014fy}		& 0.005\\
Electron spin \\
\hspace{0.5 cm}\er:\CWO	& $S_4$ 		& 0.05 \cite{Bertaina:2007dm} 		& 399 \cite{Mims:1965gx}				& 12\\ 
\hspace{0.5 cm}NV in diamond			& $C_{3v}$ 		& 1.8 \cite{Balasubramanian:2009fu}	& 17 \cite{VanOort:1990jk}			& 8.2\\ 
Nuclear spin \\
\hspace{0.5 cm}$^{151}$\eu:\YSO 	& $C_{1}$ 	&  26 \cite{Arcangeli:2014dr}		& 0.1 \cite{Macfarlane:2014fy}			& 9.6 \\ 
\end{tabular}
 \end{ruledtabular}
 \end{table}
 
%


In conclusion, we introduced a memory protocol for ensembles of atoms or spins that involves two rephasing pulses to avoid producing an output in an inverted medium. The intermediate collective emission, as well as rephased spontaneous emission are cancelled  by a Stark induced linear phase shift of centers related by an inversion symmetry. The protocol is thus low-noise and suitable for a quantum memory. Moreover, large storage bandwidths are possible since  the cancellation process is insensitive to inhomogeneities in Stark coefficients or  the electric field.  The protocol has been investigated in RE nuclear spins in a single crystal, where we found a strong echo suppression of $1.5\times 10^{-5}$. Opposite Stark phase shifts   are produced to recover the memory output, which ensures a high fidelity, experimentally confirmed by quantum state tomography. SEMM could be used to store optical or microwave photons with high efficiency in atoms and/or spin transitions coupled to cavities. This includes NV centers in diamond and rare earth doped crystals, which are currently among the most promising solid-state quantum memories.

We thank Roger Macfarlane for inspiring this study and Thierry Chaneli\`ere, Mikael Afzelius, Klaus M{\o}lmer and John Bartholomew for useful discussions and comments. This work received funding from the European Union's Seventh Framework Programme (project No. 287252, CIPRIS), ANR project DISCRYS (No. 14-CE26-0037-01), Idex no. ANR-10-IDEX-0001-02 PSL and Nano'K project RECTUS.


%

\end{document}